# The Collective Coordinates Jacobian

**Moshe Schwartz, Guy Vinograd**

*Raymond and Beverly Sackler Faculty of Exact Sciences*

*School of Physics and Astronomy*

*Tel Aviv University, Ramat Aviv, 69978, Israel*

We develop an expansion for the Jacobian of the transformation from particle coordinates to collective coordinates. As a demonstration, we use the lowest order of the expansion in conjunction with a variational principle to obtain the Percus Yevick equation for a monodisperse hard sphere system and the Lebowitz equations for a polydisperse hard sphere system.

The theory of quantum or classical fluids has developed in two main directions. One line of approach is based on the single particle concept, leading to various forms of low density expansions (cluster expansion, etc). The other direction is based on the use of the fourier components of the density as collective coordinates (RPA, etc). A major contribution to the collective coordinates approach is the work of Percus and Yevick [1] that resulted in the celebrated Percus Yevick equation for the pair distribution function of a classical fluid. There are a number of reasons for the importance of the PY equation. Firstly, it enables one to treat violent interactions, like the hard sphere potential, where the ordinary RPA fails. It produces structure factors that are quite similar to those obtained experimentally and by simulations [2], [3]. Most importantly, it is relatively simple. In fact, the hard sphere PY equation is soluble analytically [4]. An important extension of the PY equation is the set of coupled equations proposed by Lebowitz [5] that deal with a polydisperse system of hard spheres.

The purpose of the present article is to present a systematic way for describing classical equilibrium statistical mechanics of a set of interacting particles as a field theory in the density. This systematic way is based on the expansion of the Jacobian of the transformation from particle coordinates to collective coordinates. We expect that this expansion will yield powerful methods to deal with classical many body systems. In order to demonstrate its usefulness we will show that even its lowest order is of interest. We will use it in conjunction with a simple variational principle to produce extremely simple derivations of the PY and Lebowitz equations for hard sphere systems.

We consider a system of indistinguishable particles. For such a system, the only relevant physical observables are functionals of the density, $\rho(\mathbf{r})$, that is expressed in terms of the particle coordinates

$$\rho(\mathbf{r}) \equiv \sum_i \delta(\mathbf{r} - \mathbf{r}_i) . \qquad (1)$$

The explicit purpose in constructing a field theory in the density is to obtain a "free energy functional", H, (which is actually a Hamiltonian) expressed is terms of collective coordinates such, that the thermal average of any functional of the density, $G\{\rho\}$, will be given by

$$\langle G\{\rho\}\rangle = \frac{\int G\{\rho\}\exp[-\beta H\{\rho\}]D\rho}{\int \exp[-\beta H\{\rho\}]D\rho}, \qquad (2)$$

where $D\rho$ denotes functional integration. This should be distinguished from mean field calculations in which the configuration, $\rho(\mathbf{r})$, that minimizes $H\{\rho\}$ is considered to be dominant. The interested reader may already think that we are trying to reinvent the wheel ,because whether the full functional integration is carried out or just the corresponding mean field treatment is done, such a field theory in the density exists already for a long time and was exploited in numerous publications ( [6] and references there in for example). That field theory consists of an explicit form of the "free energy functional" (Hamiltonian),

$$H\{\rho\} = U\{\rho\} - TS\{\rho\}, \qquad (3)$$

where $U\{\rho\}$ is just the exact potential energy for a system of identical interacting particles:

$$U\{\rho\} = \frac{1}{2}\int d^3r d^3r' \rho(\mathbf{r})V(\mathbf{r}-\mathbf{r}')\rho(\mathbf{r}'), \qquad (4)$$

$V(\mathbf{r}-\mathbf{r}')$ being the two body interparticle potential and the "entropic contribution" is given by

$$S\{\rho\} = -k_B\left[\int d^3r \rho(\mathbf{r})\ln[\rho(\mathbf{r})/\rho_0] - \int d^3r \rho(\mathbf{r})\right], \qquad (5)$$

where $\rho_0$ is one over the cube of the thermal wave length (note that the last integral on the R.H.S of eq.(5) is not relevant when dealing with a system that consists of a fixed number of particles. For this reason, $\rho_0$ may be replaced by any other constant having the dimension of density).

The trouble is that the form of $S\{\rho\}$ given above is correct only for density fields, $\rho(\mathbf{r})$, that vary slowly. It is instructive to explain why, and to give a quantitative estimate for the slowness of variation that justifies the use of the above $S\{\rho\}$. Assume that our system contains particles of average density $\bar{\rho} \equiv \frac{N}{V}$ enclosed in a box of volume V and periodic boundary conditions apply. We represent the density field using Fourier transforms:

$$\rho(\mathbf{r}) \equiv \bar{\rho} + \frac{\bar{\rho}}{\sqrt{N}}\sum_{\mathbf{q}\neq 0}\rho_{\mathbf{q}}\exp[i\mathbf{q}.\mathbf{r}]. \qquad (6)$$

Limiting the variation of $\rho(\mathbf{r})$ means cutting off the sum on the right hand side of eq. (6) at some $q_0$. This means that the volume element over which $\rho(\mathbf{r})$ is effectively constant is of the order of $q_0^{-3}$. By noting that the derivation of eq.(5) involves the use of Sterling's formula for M!, where M is the number of particles in a volume element it is clear that this derivation is justified only for density fields that are cut off at $q_0$ such that

$$\bar{\rho} q_0^{-3} >> 1. \tag{7}$$

(Other derivations of the "enthropic contribution" rely on a steepest descent approximation, but a careful examination shows that the condition for its validity is not N>>1, but rather (7) above.)

When studying systems with short-range strong repulsive interactions, condition (7) poses an insurmountable difficulty. Assume that we are interested in the radial distribution function. The interesting structure of that function occurs at distances that are of the order of the range of the inter particle potential, R, or less. Therefore, in order to get an adequate description, we must consider density fields, $\rho(\mathbf{r})$, whose cutoffs $q_0$ are at least of order $R^{-1}$. Now, condition (7) that defines the region of validity of the form (5) for $S\{\rho\}$ reads

$$\bar{\rho} R^3 >> 1, \tag{8}$$

which is entirely impossible for hard spheres and implies extremely (unacceptable) high densities $\bar{\rho}$ for other potential models. Note that this kind of considerations must be taken into account also in other problems such as obtaining the density profile near a wall where the particles interact strongly with that wall, etc.

Here we develop a density field theory that does not have to be restricted by (7). We define now the dimensionless Fourier transform of the density (1)

$$\rho_\mathbf{q} = N^{-1/2} \sum_i \exp[-i\mathbf{q}.\mathbf{r}_i]. \tag{9}$$

Consequently, we can write the numerator, N, in eq.(2) (with a corresponding expression for the denominator) in the form:

$$N = \int d^3 r_1 ... d^3 r_N G\{\rho\} \exp\left[-\tfrac{1}{2}\beta \int d^3 r d^3 r' \rho(\mathbf{r}) V(\mathbf{r}-\mathbf{r}') \rho(\mathbf{r}')\right]$$
$$\times \prod_{\mathbf{q}\neq 0} \delta\left(\rho_\mathbf{q} - N^{-1/2} \sum_i \exp[-i\mathbf{q}.\mathbf{r}_i]\right) \prod_{\mathbf{q}\neq 0} d\rho_\mathbf{q}. \tag{10}$$

It is easy to deduce from eq.(10) that $S\{\rho\}$ given in eq.(5) has to be replaced in the "free energy functional" by

$$\tilde{S}\{\rho\} = k_B \ln \int d^3 r_1 ... d^3 r_N \rho_0^{3N} \prod_{\mathbf{q}\neq 0} \delta\left(\rho_\mathbf{q} - N^{-1/2} \sum_i \exp[-i\mathbf{q}.\mathbf{r}_i]\right). \tag{11}$$

The configurational integral in the above equation is the Jacobian, $J\{\rho\}$, of the transformation from particle coordinates, $\{\mathbf{r}_i\}$, to the collective coordinates, $\{\rho_\mathbf{q}\}$. It is clear that the Jacobian can be written in the form

$$J\{\rho\} = \rho_0^{3N} V^N \left\langle \prod_{\mathbf{q}\neq 0} \delta\left(\rho_\mathbf{q} - N^{-1/2} \sum_i \exp[-i\mathbf{q}.\mathbf{r}_i]\right) \right\rangle_I, \tag{12}$$

where $\langle \rangle_I$ denotes an ideal gas average. Using the Fourier representation of the $\delta$ function, we find that

$$J\{\rho\} \propto \int \prod_{\mathbf{q}\neq 0}(dk_\mathbf{q}(2\pi)^{-1}\exp[i\rho_\mathbf{q} k_{-\mathbf{q}}])\left\langle \exp[-iN^{-1/2}\sum_{\mathbf{p}\neq 0}k_\mathbf{p}\sum_i \exp[i\mathbf{p}.\mathbf{r}_i]]\right\rangle_I. \quad (13)$$

By using the cummulant expansion for the average and calculating some ideal gas correlations, we find that the logarithm of the average in the equation above, to be denoted $\hat{H}\{k_q\}$, is given by

$$\hat{H}\{k_\mathbf{q}\} = N\sum_{n=2}^{\infty}\frac{(-i)^n}{n!N^{n/2}}\sum_{\mathbf{l}_i\neq 0, \sum \mathbf{l}_i=0}{}' k_{\mathbf{l}_1}..k_{\mathbf{l}_n}, \quad (14)$$

where $\sum'$ implies that the set of indices, $\{l_i\}$, does not have subsets that sum up to zero (Each term in the sum over n should be of order 1 (note the N in front) and expression (14) above is correct to that order but has corrections that are higher orders in $N^{-1}$.)

The Jacobian of the transformation can now be written formally in a way that enables a systematic expansion: by denoting $\Delta\hat{H}\{k_q\} \equiv \hat{H}\{k_q\} + \frac{1}{2}\sum_{\mathbf{q}\neq 0}k_\mathbf{q} k_{-\mathbf{q}}$, it is straightforward to show that

$$J\{\rho_\mathbf{q}\} = \exp\left[\Delta\hat{H}\left\{-i\frac{\partial}{\partial \rho_{-\mathbf{q}}}\right\}\right]\exp[-\frac{1}{2}\sum_{\mathbf{q}\neq 0}\rho_\mathbf{q}\rho_{-\mathbf{q}}]. \quad (15)$$

The full Jacobian should give a full description of the ideal gas, namely to produce correctly all the density correlations of the ideal gas. The usefulness of the expansion can be checked by considering to what extent are really the ideal gas correlation produced by low orders of that expansion. We find that lowest order yields the correct two body correlations, $\langle \rho_\mathbf{q}\rho_{-\mathbf{q}}\rangle$, but not the higher order ones. When going to first order we find that $J\{\rho_\mathbf{q}\} = \left[1+(6\sqrt{N})^{-1}\sum_{\mathbf{k},\mathbf{l}}\rho_\mathbf{k}\rho_\mathbf{l}\rho_{-\mathbf{k}-\mathbf{l}}\right]\exp[-\frac{1}{2}\sum_{\mathbf{q}\neq 0}\rho_\mathbf{q}\rho_{-\mathbf{q}}]$. We see that the correct two body correlations are still produced and in addition, we obtain the correct three-body correlations. Second order expansion adds the correct four-body correlations. It is interesting that when expanding the "$\rho\ln\rho$" term in eq.(5) in $\delta\rho(\mathbf{r})/\overline{\rho}$, not only the lowest order coincides with our lowest order [7], but also the next order does. The first deviations appear in the second order, producing wrong four-body correlations in the ideal gas.

We will demonstrate next the usefulness of our expansion by deriving the PY equation for mono-disperse hard sphere system and the Lebowitz equations for polydisperse hard sphere system. This will be done by considering the zeroth order of the expansion of the Jacobian.

In terms of collective coordinates we can now define a Hamiltonian

$$H \equiv -\beta^{-1} \ln J\{\rho_{\mathbf{q}}\} + \frac{1}{2}\sum_{\mathbf{q}\neq 0} \bar{\rho} V(\mathbf{q})\rho_{\mathbf{q}}\rho_{-\mathbf{q}} \equiv H_I + H_{int}, \qquad (16)$$

where $V(\mathbf{q})$ is the Fourier transform of the inter-particle interaction and where the factor $\bar{\rho}$ comes from our definition (6) of $\rho_{\mathbf{q}}$ that makes it dimensionless.

Once H is given in terms of the collective coordinates, $\{\rho_{\mathbf{q}}\}$, a variational approach to the calculation of the structure factor, $S_{\mathbf{q}} = \langle \rho_{\mathbf{q}}\rho_{-\mathbf{q}}\rangle$, follows easily. We write

$$H\{\rho_{\mathbf{q}}\} = \frac{1}{2}\beta \sum_{\mathbf{q}\neq 0} \rho_{\mathbf{q}} S_{\mathbf{q}}^{-1} \rho_{-\mathbf{q}} + H\{\rho_{\mathbf{q}}\} - \frac{1}{2}\beta \sum_{\mathbf{q}\neq 0} \rho_{\mathbf{q}} S_{\mathbf{q}}^{-1} \rho_{-\mathbf{q}} \equiv H_0 + H - H_0. \qquad (17)$$

It is well known that the correct free energy, F, obeys an exact inequality:

$$F \leq F^*, \qquad (18)$$

where

$$F^* = F_0 + \langle H - H_0 \rangle_0, \qquad (19)$$

$F_0$ being the free energy resulting from $H_0$ and $\langle ... \rangle_0$ denotes thermal average with respect to $H_0$. The unknown structure factor, $S_{\mathbf{q}}$, is obtained now by minimization of $F^*$ with respect to $S_{\mathbf{q}}$. This method does not work for a hard sphere system because the inter-particle potential does not have a Fourier transform. The hard sphere system can be viewed as an ideal gas and an infinite number of constraints on the correlation functions. For distinct positions, $\{\mathbf{r}_i\}$, we must have $\langle \rho(\mathbf{r}_1)\rho(\mathbf{r}_2)...\rho(\mathbf{r}_n)\rangle = 0$ for all $n \geq 2$ if there exists a pair $(i,j) \in \{1,2,...,n\}$ such that $|\mathbf{r}_i - \mathbf{r}_j| \leq R$, where R is the diameter of the hard sphere. The minimization of $F^*$ with respect to the $S_{\mathbf{q}}$'s will have to be performed subject to the above constraints. We replace now the full ideal gas Hamiltonian, $H_I$, by its lowest order approximation

$$\tilde{H}\{\rho_{\mathbf{q}}\} = \frac{1}{2}\beta^{-1}\sum_{\mathbf{q}\neq 0}\rho_{\mathbf{q}}\rho_{-\mathbf{q}}, \qquad (20)$$

and relax the hard sphere constrains by imposing only the two body constraint on the pair distribution function

$$g_2(\mathbf{r}) \equiv \frac{1}{\bar{\rho}^2}(\langle \rho(\mathbf{r})\rho(0)\rangle - \langle \rho(0)\rangle \delta(\mathbf{r})), \qquad (21)$$

that is required to obey

$$g_2(\mathbf{r}) = 0, \qquad (22)$$

for all $r \leq R$. This is the constraint used by PY [4].

From the structure factor definition it follows that the Fourier presentation of $g_2(\mathbf{r})$ is :

$$g_2(\mathbf{r}) = 1 + N^{-1} \sum_{\mathbf{q} \neq 0} (S_\mathbf{q} - 1) \exp[i\mathbf{q}.\mathbf{r}]. \qquad (23)$$

The functional $F^*$ is easily calculated to be

$$F^* = \frac{1}{2} \sum_{\mathbf{q} \neq 0} (-\ln S_\mathbf{q} + S_\mathbf{q} - 1). \qquad (24)$$

It has to be minimized with respect to $S_\mathbf{q}$ with the constraint (22) (that is in fact an infinite number of constraints holding for all $|\mathbf{r}| < R$). These are introduced via incorporating Lagrange multipliers $\lambda(\mathbf{r})$ for all $|\mathbf{r}| < R$. The minimization results in

$$0 = (1 - S_\mathbf{q}^{-1}) - \int_{|\mathbf{r}|<R} d^3 r \lambda(\mathbf{r}) \exp[i\mathbf{q}.\mathbf{r}] \equiv (1 - S_\mathbf{q}^{-1}) - \overline{\rho \tilde{v}}(\mathbf{q}), \qquad (25)$$

where $\tilde{v}(\mathbf{q})$ is a Fourier transform of a function of the absolute value of $\mathbf{r}$ ($\lambda(\mathbf{r})$) that vanishes for r>R. We see from (25) that

$$S_\mathbf{q} = (1 - \overline{\rho \tilde{v}}(\mathbf{q}))^{-1}, \qquad (26)$$

so that $\overline{\rho \tilde{v}}(\mathbf{q})$ is to be identified with $C(\mathbf{q})$, the Fourier transform of the Orenstein-Zernike [8] direct pair correlation function, $C(\mathbf{r})$. Thus our equations are

$$g_2(\mathbf{r}) = 0 \text{ for } r \leq R \text{ and } C(\mathbf{r}) = 0 \text{ for r>R}, \qquad (27)$$

which are the PY equation [4] for a system of hard spheres.

For a system that consists of different types of identical particles, the Jacobian of the transformation from particle to collective coordinates is just the product of the Jacobians for each type separately. The number of density fields is the number of the different types, of course. In order to obtain the Lebowitz equations [5] for a mixture of hard spheres, we have to note that a particle of type "i" and a particle of type "j" cannot be within a distance less than the corresponding hard sphere radius, $R_{ij}$. This yields a natural extension of condition (22) that together with the above variational principle leads easily to the Lebowitz equations [5].

[1]   J.K. Percus and G.J. Yevick, Physical Review, **110**, 1 (1958)